\documentclass[12pt,manuscript]{aastex}
\usepackage{color}
\usepackage{ulem}
\usepackage{lineno}

\shorttitle{[Sr/Ba] in Metal-Poor Stars}
\shortauthors{Aoki et al.}

\begin{document}
%\linenumbers
%% LaTeX will automatically break titles if they run longer than
%% one line. However, you may use \\ to force a line break if
%% you desire.

\title{Explaining the Sr and Ba Scatter in Extremely Metal-Poor Stars}
\author{W. Aoki}
\affil{National Astronomical 
Observatory of Japan, 
2-21-1 Mitaka, Tokyo, 181-8588, Japan}
\email{aoki.wako@nao.ac.jp}
\author{T. Suda}
\affil{National Astronomical 
Observatory of Japan, 
2-21-1 Mitaka, Tokyo, 181-8588, Japan}
\email{takuma.suda@nao.ac.jp}
\author{R.N. Boyd}
\affil{National Astronomical 
Observatory of Japan, 
2-21-1 Mitaka, Tokyo, 181-8588, Japan}
\email{richard11boyde@comcast.net}
\author{T. Kajino\altaffilmark{1}}
\affil{National Astronomical 
Observatory of Japan, 
2-21-1 Mitaka, Tokyo, 181-8588, Japan}
\email{kajino@nao.ac.jp}
\and
\author{M.A. Famiano\altaffilmark{2}}
\affil{National Astronomical 
Observatory of Japan, 
2-21-1 Mitaka, Tokyo, 181-8588, Japan}
\email{michael.famiano@wmich.edu}
\altaffiltext{1}{Dept. of Astronomy, 
Graduate School of Science; Univ. of Tokyo, 7-3-1, 
Hongo, Bunkyo-ku, Tokyo 113-0033, Japan}
\altaffiltext{2}{Dept. of Physics and Joint Institute for Nuclear Astrophysics, Western Michigan University, 
1903 W. Michigan Avenue, Kalamazoo, MI 49008-5252, USA}
\begin{abstract}
Compilations of abundances of Strontium and 
Barium in
extremely metal-poor stars show that an apparent cutoff is observed for
[Sr/Ba] at [Fe/H]$<$-3.6 and large fluctuations for [Fe/H]$>$-3.6 with a 
clear upper bound depending on metallicity.
We study the factors that place upper limits on the 
logarithmic ratio
[Sr/Ba].  A model is developed in which the
collapses of type II supernovae are found to reproduce many of the features
seen in the data.  This model
is consistent with galactic chemical evolution constraints of light-element 
enrichment in metal-poor stars. Effects of turbulence in an explosive site 
have also
been simulated, and are found to be important in explaining the 
large scatter observed in the [Sr/Ba] data.

\end{abstract}
\maketitle
\section{Introduction}
Nuclei heavier than iron and nickel are mostly made either in the s-process 
(slow neutron-capture process) or r-process (rapid neutron-capture process)
\citep{b2fh, wallerstein97}, which is 
thought to produce half the nuclides heavier than Fe and nearly all actinides 
observed in most metal-poor stars.  Although effects of the s-process can 
occur in first generation stars \citep{goriely01, suda04, nishimura09},
the s-process appears not to be responsible for the enrichment of 
the interstellar medium with the elements produced in the early life of the
galaxy because it requires longer timescales to populate the ISM with its products.
The r-process, however, produces its nuclei 
earlier in the galactic history.
Because massive stars produce most of the required conditions for 
the r-process, they have long been supposed to synthesize its nuclides. 
However, detailed calculations \citep{fischer10, hudepohl10, goriely11}
have suggested that massive stars may not fulfill all of the r-process conditions. 
We present data that 
sheds important new light on this question.  

Stars with masses greater than 8 $M_\odot$ are 
thought to become 
core-collapse supernovae. Those less than 25 $M_\odot$ will 
form neutron stars, emitting neutrinos and driving some 
of their newly synthesized nuclei into the interstellar medium. 
Those more massive than about 40 $M_\odot$ are thought to 
collapse directly to black holes, 
contributing nothing to the interstellar medium. Stars with 
25$\le M \le$ 40 $M_\odot$ are thought to collapse to neutron stars then, 
as some of the expelled material falls back onto the neutron 
star, to black holes \citep{fryer03, heger92},
producing ``accretion-induced black holes'' 
and, assuming some material can escape, 
would be expected to contribute some newly synthesized nuclides to the 
interstellar medium.

We report on some striking astronomical data from extremely 
metal-poor (EMP) stars, those having [Fe/H]$<$-2.5.  Their metals are thought to 
have been produced
in the first generation massive stars that were the progenitors of 
the ones currently observed. 
Because these
stars have very low abundances of anything except primordial hydrogen and helium, 
their abundances of heavy nuclei are presumed to arise primarily from the r-process.
We concentrate on elements made in the r-process and that are easy to detect 
in these stars: strontium and barium.  
%A parametrized galactic chemical evolution model is presented
%by which the production of these elements in metal-poor stars is explained via the r-process that 
%occurs in a single supernova; its results are compared to sites in which multiple r-process events
%are described.

Previous models \citep{qw,wanajo,LEPP} have described separate r-process sites
as the cause of an enrichment of light neutron-capture r-process elements.  
We contrast our model with the ``weak'' r-process and the light element 
primary process (LEPP) by describing an enrichment in the lighter neutron-capture elements
\citep{honda}
that is a natural consequence of a single site but with differing levels of enrichment of light 
and heavier neutron-capture 
elements.  
Our model produces the r-process from an
explosive event (e.g., a type II supernova), 
while the r-process ejecta is prohibited in some fraction of those events 
from escaping due to a subsequent collapse of the
proto-neutron star to a black hole.  The 
collapse time in this event relative to the r-process ejection time will determine how much
r-process material is ejected, as well as its composition.
%We note that there are significant similarities between our 
%model and the phenomenological three component model of \citet{qw}.
%The QW model has two of their three components competing in the 
%EMP star region, with hypernovae dominating for [Fe/H] < -3.6. 
%They attribute production only of lighter nuclei, especially Fe, to 
%hypernovae (although other studies \citep{nakamura11, caballero12} 
%suggest that hypernovae might also produce 
%r-process nuclides). The QW model explains the large dispersion of [Sr/Ba] 
%data by assuming spatial inhomogeneities in the products of the two 
%competing components, and assuming a parameter to define the relative 
%contributions of the two components. 
%
%Mathematically the QW model apears to be similar to our model, although our model 
%arises more from the assumed
%sources of the observed nucleosynthesis than 
%does the QW model. The mixing parameters of QW could be used to characterize 
%the relative amounts of Sr (r-process seeds) and Ba (r-process nuclides) that 
%the turbulence of the fallback supernovae were able to expel into the 
%interstellar medium. The two models differ primarily in the sharpness of the 
%cutoff at [Fe/H] = -3.6; it would be expected to be more gradual in 
%the QW model than in our model. Unfortunately, the present data set of 
%EMP stars is insufficient to discriminate between the two models; it may
%require many years of data taking on EMP stars before that can be achieved.
\section{Observational Data}
Data were taken from the SAGA Database \citep{saga}. The
sample includes data from 186 papers and 4270 stars (1488 unique
stars) dealing with observations of
metal-poor halo stars published prior to 2012. 
The number of stars for
which [Fe/H]$<$-2.5 is 428, reducing to 361 stars if carbon-enhanced EMP stars
are excluded.
This sample could suffer errors due to systematic
offsets between different studies; a comparison of results for the same
stars but from different studies shows that these can sometimes be as large as
a factor of 3. We thus selected
the results of Sr and Ba from observations in which both elements are
reported in the same paper - 260 total stars. 
The 
abundances in this database were all
obtained assuming local thermodynamic equilibrium (LTE). The scatter
found in Sr/Ba as well as in Sr/Fe and Ba/Fe in that sample in the
metallicity range of -3.6$<$[Fe/H]$<$-3.0 
is observed to be a factor of 100 or more, much larger
than the observational errors.

Figure \ref{sr_ba_fe_h} shows the astronomical data for [Sr/Fe], [Ba/Fe], and [Sr/Ba]
from the data set described above, showing  a wide
range of [Sr/Ba], [Sr/Fe], and [Ba/Fe] values at
each value of [Fe/H]. Our interpretation of these features will be
given in subsequent sections; the curves in Figure \ref{sr_ba_fe_h} that overlay the
data, reflect these efforts.

We find three remarkable features in the [Sr/Ba] distribution.
\begin{enumerate}
\item Almost all stars appear in the range of the functional
[Sr/Ba]$<-1.5-$[Fe/H].
There are two exceptions in [Fe/H]$<-2.0$. One of them is
BD+80$^o$245 ([Fe/H]$=-2.09$, [Sr/Ba]$=+1.04$), which is a well
known "$\alpha$-deficient" star (e.g., [Mg/Fe]$=-0.22$) having
exceptionally low abundances of neutron-capture elements
([Sr/Fe]$=-0.85$ and [Ba/Fe]$=-1.89$ which was studied by \cite{ivans}). The other is
HE0029-1839 studied by \cite{barklem}, for which little 
information exists. 
\item There are only a few stars with
[Sr/Ba]$<$-0.5. We present calculations that show that this 
range is not expected from the r-process or from
tr-process, which we discuss in the next section. 
The exceptions are as follows: 
\begin{itemize}
\item CS~30322--023 ([Fe/H]=-3.25, [Sr/Ba]=-1.10): a moderately
  carbon-enhanced star ([C/Fe]=+0.6) with enhancement of nitrogen and
  Ba \citep{masseron,aoki07}.
\item CS~29493--090 ([Fe/H]=-2.82, [Sr/Ba]=-1.41): a moderately
carbon-enhanced star ([C/Fe]=+0.74) with excess of Ba \citep{barklem}.
\item CS~22946--011 ([Fe/H]=-2.59, [Sr/Ba]=-0.99) and CS~22941--005
  ([Fe/H]=-2.43, [Sr/Ba]=-0.77): binaries with excesses of Ba
  ([Ba/Fe]=+1.26 and +0.34, respectively: \cite{preston}).
  CS~22950--173 ([Fe/H]=-2.5, [Sr/Ba]=-0.72) which also belongs to be a binary
  system, though Ba is not enhanced ([Ba/Fe]=-0.04:
  \cite{preston}). The CH feature would not be detected due to high
  temperature ($\sim$6800K).
\item HE0305-4520 ([Fe/H]=-2.91, [Sr/Ba]=-1.25): Ba is slightly enhanced
([Ba/Fe]=+0.59), while an excess of carbon is not clear ([C/Fe]=+0.33:
\cite{barklem}).
\end{itemize}
Hence, except for HE0305-4520, all the above stars are (candidate)
carbon-enhanced stars, 
which can be affected by an s-process contribution through
mass transfer in binary systems.
HE0305-4520 could
also be slightly affected by s-process. In the objects with low Sr
abundances, Sr/Ba could be driven to very low values even by a small contamination
of Ba from the s-process.

There is no object 
for which s-process contamination is not suspected
in the region [Sr/Ba]$<-0.5$ given the typical
errors ($\sim$0.2 dex or larger) in [Sr/Ba]. 

\item A cutoff in [Sr/Ba] data at 
[Fe/H]$\sim -3.6$, which will be addressed below, is found. There
 are only four stars (CS~22172--002, CD$-38^\circ$245, CS~22885--296, and CS~30336--049) 
in [Fe/H]$<-3.6$ for which both of the Sr and Ba
abundances are determined, and all of them have low [Sr/Ba]
($<0.05$).
There are roughly three times as many stars with [Sr/Ba]$>0.05$ than
with [Sr/Ba]$<0.05$ in the region
$-3.6 < $[Fe/H]$ < -3.0$.
If the same distribution is assumed for $-4.0 < $[Fe/H]$ < -3.6$,
one would expect 12 - 15 stars with [Sr/Ba]$>$0.05, instead of the zero
stars that are observed, from the four stars with [Sr/Ba]$>$0.05.

To confirm this cutoff, we inspected the data with [Fe/H]$ <
-3.6$ in more detail. We found 18 objects in this range in our sample, but
both Sr and Ba abundances were determined only for seven objects, among
which three are carbon-enhanced objects. 

Among the remaining 11 objects for which [Fe/H]$ < -3.6$, five stars
are carbon-enhanced, and four have very low [Sr/Fe] or low upper
limits, suggesting that Sr/Ba ratios are also quite low. There is no
signature to have high Sr/Ba for other two objects although the upper
limits for Ba abundances are still weak. 

We note that no new stars with both Sr and Ba measurements for
[Fe/H]$<-3.6$ were reported by the recent paper to study chemical
composition of most metal-poor stars by \citet{yong12}.
\end{enumerate}
\section{Galactic Chemical Evolution Interpretation and Turbulence}
The most daunting challenges for any attempt to 
explain the data distribution 
for the EMP stars are reproduction of the sharp cutoff in [Sr/Ba]
at [Fe/H]=-3.6, 
the explanation of the upper and lower limits as a function of metallicity 
in that distribution, and the wide dispersion in [Sr/Ba].

A reduction in the Sr and Ba production
in a SN collapse scenario was studied previously \citep{bfak}, 
using the formalism and results from \cite{woo94}.
There, the r-process occurs in upper-most 15 shells that appear just at the surface of the 
nascent neutron star.  The mass, thermodynamic trajectories, and initial
composition of each shell are described in \cite{woo94}.
Those shells were examined in the framework of the
ejection time, the black-hole collapse time, and the resultant nucleosynthesis, assuming
that the ejection is halted by the collapse to a black hole.  We assume that any collapse time occurring prior
to the shell ejection time in this model will prevent that shell from being ejected. 
At present, we extend that model to study the
relationship between the stellar metallicity, the progenitor mass, the
collapse time, and elemental yields.  This model is coupled to a 
galactic chemical evolution (GCE) code \citep{timmes}.  The
net result is that a more massive progenitor will generally collapse
earlier, resulting in ejecta enriched
in the lighter neutron-capture elements (A$\lesssim$130) but a lower yield of the heavier
elements.  This GCE formulation will be described in detail in a subsequent 
paper \citep{fbaks}.  Here, we present the general principles.

Previous results have indicated that a truncation in the r-process in
type II supernovae due to a collapse of the proto-neutron star to a 
black hole may be responsible for a relative enrichment of 
light neutron-capture elements in some metal-poor
halo stars produced by a single-site ``truncated r-process'' \citep{bfak} (where
single-site is defined to be a single episode of star formation after one
r-process event).  This 
``tr-process'' could be caused by either a dynamic collapse of
material below the black hole mass cut or by a change in the neutrino
luminosity resulting in an increased electron fraction $Y_e$ in the later stages of the
r-process, which would result in a reduced  synthesis of the r-process nuclei with masses in excess 
of A$\sim$130.

We adopt a model based on prior calculations \citep{bfak} in which the isotopic
yields in a tr-process are directly related to the collapse time in an
accretion-induced black hole.  
In order to produce the chemical evolution results shown, 
we began with the Ba yields of \cite{travaglio} and performed several calculations 
iteratively to arrive at the yields shown in Figure \ref{yields}.  These assumed
yields reproduce the GCE results shown by the red lines in Figure \ref{sr_ba_fe_h}.
For stars with 10 M$_\odot <$M$<$ 30 M$_\odot$, the yields are 
roughtly equal to those of Model 1Max in \cite{travaglio} but slightly less than those 
of model 2Max in the same reference and less than
those of \cite{raiteri} by about a factor of four (for 10-11 M$_\odot$ SNIIs).  
For M$<$10 M$_\odot$ yields are estimated from Figure 3 of \cite{travaglio}.
The Sr yields (Figure \ref{yields}) were chosen to reproduce the [Sr/Fe] and [Sr/Ba] GCE results in Figure \ref{sr_ba_fe_h}.  The only significant difference between the Ba yields in this paper 
and those of \cite{travaglio} corresponds to 30 M$_\odot$ stars, 
which are roughly 30 times higher 
than those of model 1Max.  This produces
Ba early in the GCE model, though the contribution is insignificant at late times.

A power law IMF was used with a 
Salpeter exponent 
of -1.31.  
The black hole collapse time
was computed based on progenitor mass and metallicity and the nuclear
equation of state (EOS) \citep{GRID_ref}.  
This 
time was then assumed to be the truncation time in a dynamic tr-process.  For
later truncation times, mass shells closer to the proto-neutron star core, blown off
later in the
explosion model, are assumed to be ejected either by achieving a velocity in excess of the
escape velocity or perhaps by being ejected in a jet. 
Later collapse
times correspond to larger production yields of heavier r-process elements. 
%The mass-weighted 
%yields of Sr and Ba
%produced in a tr-process can be scaled based on the collapse time, which ultimately depends on the 
%progenitor mass and metallicity \citep{fbaks}.

In the r-process, most of the Sr is produced in the early-stage ejectt in mass shells
ejected within the first four seconds post-bounce, while most of the
Ba is produced in the later-stage ejecta of a SNII in mass shells
ejected from between four and 18 seconds post-bounce.  Thus, the Sr yields in a tr-process
are fairly constant with collapse time and fairly close to their complete r-process values
except for very short collapse times for which no r-process material is ejected.  
As only fairly late collapse times
will allow for a significant amount of Ba ejection in the tr-process, [Sr/Ba] approaches zero in the later-stage ejecta. 
%The Ba
%yields, however, do not
%reach their full r-process values until fairly late collapse times (several seconds post-bounce).

Results from the 
spherical collapse code GR1D\citep{GRID_ref} 
were used to study the relationship between the progenitor star's mass
and metallicity and the Sr and Ba yields in massive stars.  
%This involves the black hole collapse time, for
%several nuclear EOSs.  
It is noted that these yields are minimum yields, since
in the non-rotating spherical models employed here, the collapse times
are minimum collapse times.  Longer collapse times may 
result from effects such as asymmetric explosions, rotations, and neutrino heating induced by these effects.  However,
since Sr is produced in shells that are ejected very early on, increasing the collapse time will have little effect
on the Sr production, while significantly increasing the Ba production.

Ejection of supernova materials into the interstellar medium resulting 
from turbulence in the stellar interior may also affect the observed abundances. 
Although it is difficult to simulate such effects in a one-dimensional 
model, we have performed a simplified analysis that at least suggests the 
sort of effects that might be observed, and probably to constrain 
the magnitude of the predictions. This analysis took each 
of the 15 individual mass shells that were used in our analysis and that 
could be ejected from the surface of the neutron star.
We then assumed that
turbulent ejection might allow a single individual shell to be 
ejected from the star, while the others collapsed
onto the proto-neutron star surface.  As noted above, 
the outer shell would be rich in Sr and would have very little Ba, 
while the inner shells would have much less Sr and would be much richer 
in Ba. Ejection of only the top or bottom shells 
should provide limits on the ability of 
turbulence to explain the range of [Sr/Fe], [Ba/Fe], and [Sr/Ba] observed
and might also explain the dispersion in the data that mixing might produce.  
The [Sr/Ba] results from sites
ejecting only these shells are shown in the GCE model in Figure \ref{sr_ba_fe_h}d.   
\section{Results}
\label{results}
Assuming a tr-process model resulting in a reduced Sr and Ba production, 
the single-site [Sr/Ba] values as a function
of metallicity [Fe/H] are shown in Figure \ref{sr_ba_fe_h} with a collapse model
utilizing the LS220 EOS \citep{ls_eos}, which is one of the most
widely used EOSs in numerical simulations of supernova explosions.  
Several other EOSs have
been examined, and it has been found
that a softer EOS results in progressively larger [Sr/Ba] values at lower 
metallicity (corresponding to earlier times in the galactic evolution produced
by stars with M$\ge$ 20 M$_\odot$) \citep{fbaks}. 

In Figure \ref{sr_ba_fe_h}, two GCE results are presented.  One is
the metallicity relationship from a model based on the input yields
from Figure \ref{yields} showing [Sr/Fe], [Ba/Fe], and [Sr/Ba] ratios.  It can
be seen that those fall within the extremes of the observational data points.  The other result
is a metallicity relationship corresponding to a GCE model in which
{\bf all} stars with M$\ge$20M$_\odot$ ultimately collapse to black holes.
% with collapse
%times dictated by a non-rotating progenitor, resulting in
%a tr-process. 

%In this latter case, the collapse time is expected to be a minimum since rotation
%will increase the collapse time.  
In this model, an earlier collapse means that shells
closer to the neutron star surface (which produce more Ba) will not be ejected, and 
the overall Ba ejection (summed over all shells) is reduced.  
Since Sr is mostly produced in outer shells, it is
not reduced as much.  The net result
is that the calculated lines in the figure will correspond to 
the minimum in the [Sr/Fe] and [Ba/Fe] 
values and a maximum in the [Sr/Ba] values.  Collapse times range from about 0.25s for 
40 M$_\odot$ stars with a soft EOS to over 3.5s for 20 M$_\odot$ stars with a stiff EOS. 
%The scatter in values from those of
%this extremely simple model would then
%be caused by increased collapse times (due to rotation, netrino heating, %and asymmetric 
%explosion mechanisms, for example) with points falling between those of %the
%standard GCE model and the tr-process GCE model.

The results of these calculations can be compared to the astronomical data of [Sr/Ba]
discussed here. These
observations show a sharp
maximum in the [Sr/Ba] value for a given metallicity. 
This could be explained by the tr-process limiting
the
Sr or Ba production as a function of stellar age or mass.  We also note a similarly 
sharp lower limit in the [Ba/Fe] values at [Ba/Fe]$<$0 corresponding to the
sharp [Sr/Ba] upper limit.  However, the lower limit in [Sr/Fe] is less 
distinct.  
%One possible explanation for this is that Ba production for 
%EMP stars is dominated by a primary r-process, while the Sr production 
%can have a significant
%s-process contribution.  
%The net result is that the maximum calculated [Sr/Ba] ratio is near the 
%upper limit in observed
%values.
%
The curves shown in Figure \ref{sr_ba_fe_h}d represent attempts to study the 
effects of turbulence by assuming that individual mass shells were 
ejected into the ISM. It can be seen that these results do span the 
entire range of observed [Sr/Ba] values, and even exceed the 
limits of the data. Obviously, in reality 
one would expect some mixing of shells to occur.
As noted above, what is 
happening qualitatively is that the outermost shell is high in Sr and 
low in Ba, so produces a very large [Sr/Ba] value, whereas the innermost 
shells are low in Sr and high in Ba, so produce the lowest [Sr/Ba]
values in the present GCE model. 
Additional smearing of these results might be expected from other 
effects that are not included in our simple calculations,
%such as 
%rotation and convection induced by
%neutrino heating, 
but the turbulence alone does appear to 
provide an explanation of 
the wide scatter observed in the data. 

If the atmospheres of these EMP stars 
are enriched by the tr-process, then the EOS
dependence of the collapse time has a direct effect on the Sr and Ba
production.  Since the plotted values of [Sr/Ba] are already the  
maximum values, it would appear that in this model 
a very stiff EOS cannot produce the large observed values of [Sr/Ba]; a
softer EOS is necessary.  It's also noted that the upper limit of
observed [Sr/Ba] values provides an observationally imposed lower limit on the
softness of the EOS.
%Thus, the EOS dependence of an accretion-induced black hole collapse 
%is shown to have a direct effect on the production of Sr and Ba
%and their subsequent abundance ratios.  
%Details will be discussed in a 
%forthcoming paper \citep{fbaks}.
\section{Conclusions}
Our detailed inspection of Sr and Ba abundances in Galactic metal-poor
stars indicates a clear cutoff in the distribution of [Sr/Ba] at [Fe/H]$
\sim -3.6$ and an upper bound as a function of [Fe/H] ([Sr/Ba]
$<-1.5-$[Fe/H]) as well as a lower bound at [Sr/Ba]$\sim -0.5$.
This work lends support to the suggestion \citep{bfak} that the cutoff 
at [Fe/H]=-3.6 for [Sr/Ba]$>$0.0 
data for EMP stars can be explained by the tr-process, that is, by the 
collapses to black holes over a 
fairly wide mass range of progenitor stars. Furthermore, the results of our GCE calculations are 
found to 
%exhibit the sharp cutoff that is seen in the observational data. These 
%results 
predict the upper bound in [Sr/Ba] in these data.
%and several mechanisms exist 
%that are not included in our simple model that would be expected to %produce the 
%large scatter that is observed in these data.

\begin{figure}
\plotone{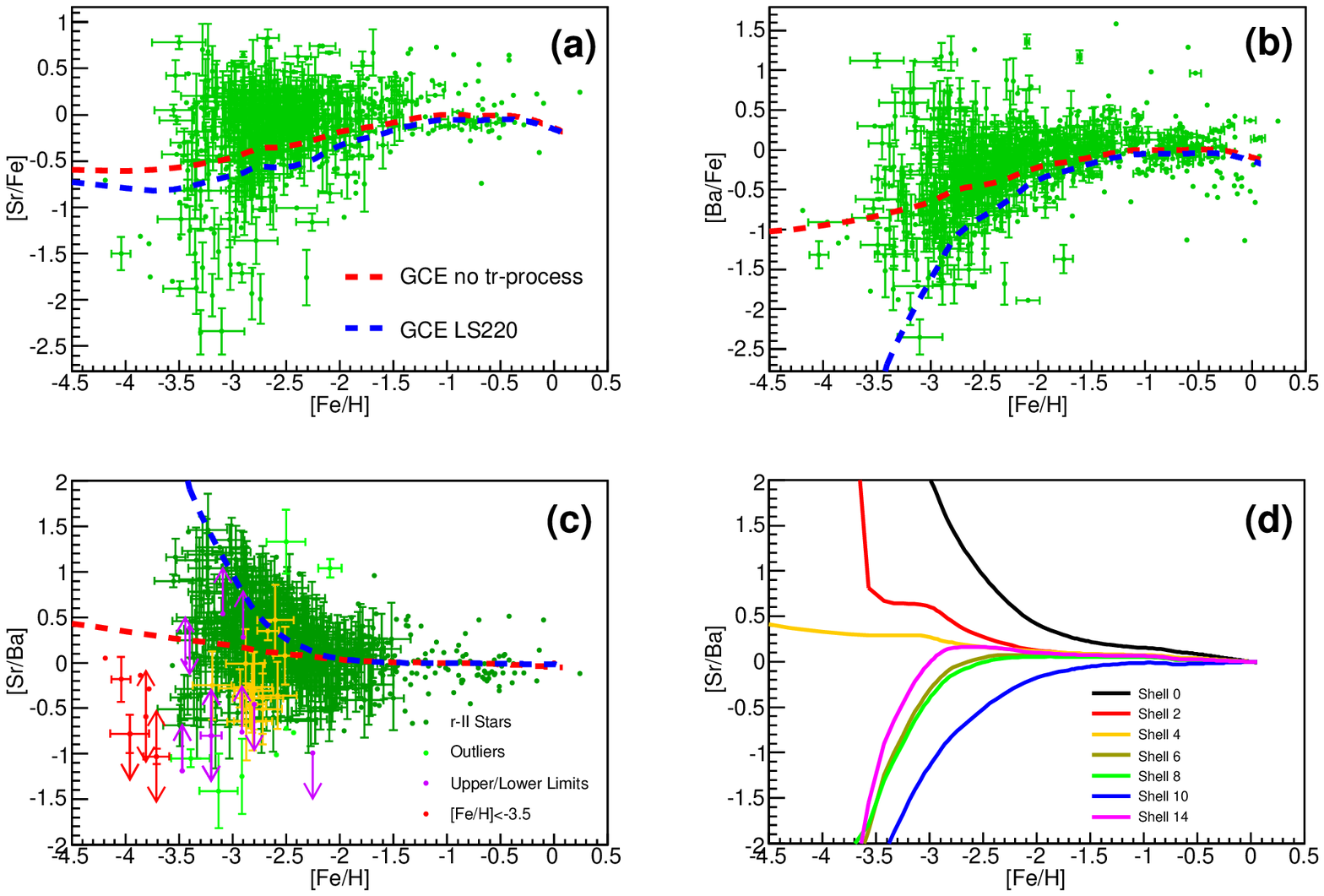}
\caption{Results for (a) [Sr/Fe], (b) [Ba/Fe], and (c) [Sr/Ba] assuming a tr-process.
The red dashed line corresponds to a GCE model with no production in a tr-process, and the blue 
dashed line corresponds to a GCE model with a primary production from a tr-process for all stars with
M$\ge$20M$_\odot$. Plot (d) shows GCE results for single-site tr-process production for turbulent ejection of
specific shells assuming only those shells are ejected.
\label{sr_ba_fe_h}}
\end{figure}

\begin{figure}
\includegraphics[scale=0.65]{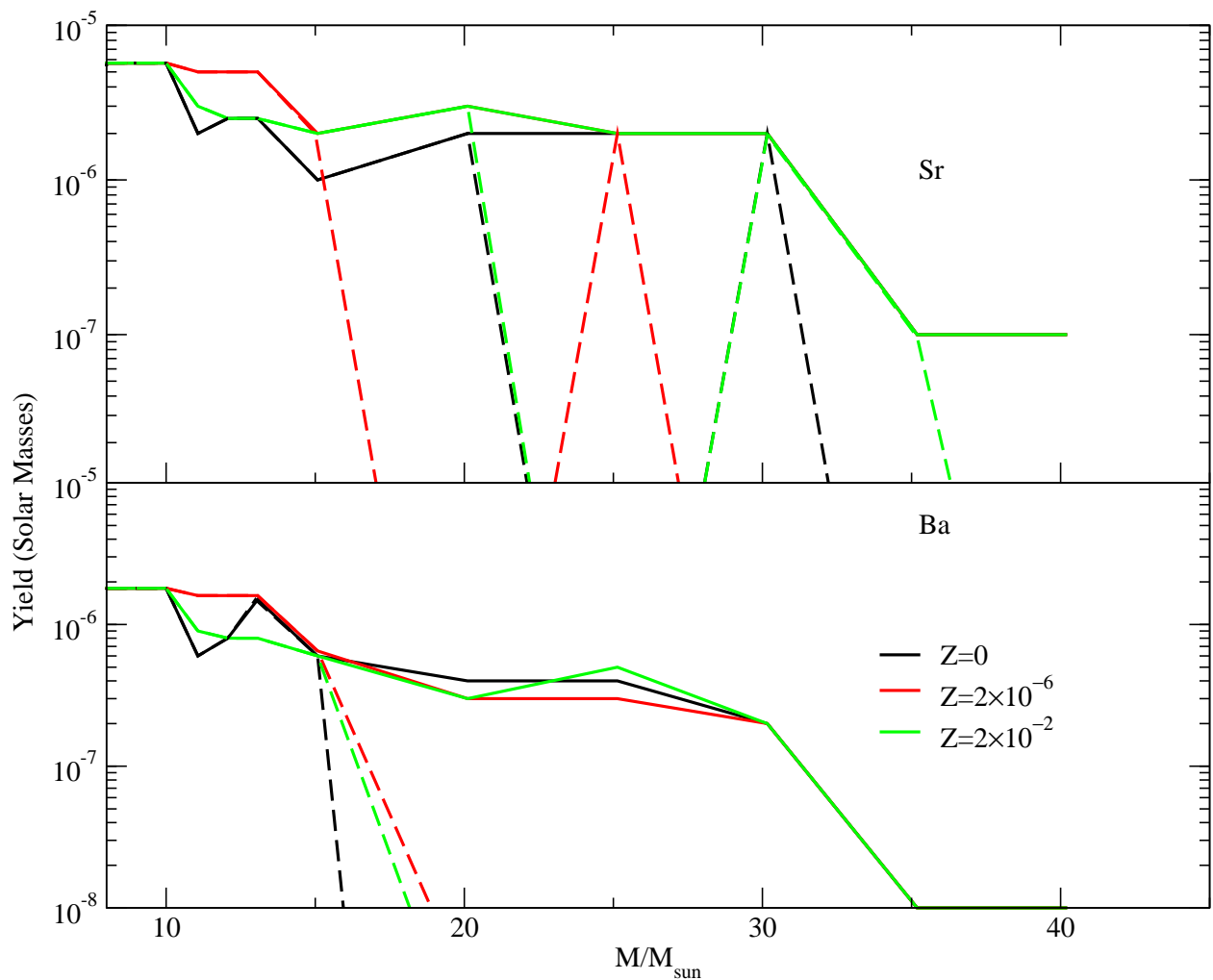}
\caption{Yields of Sr and Ba used for the GCE calculations in this model as
a function of stellar mass and metallicity.  
The solid lines correspond to yields assumed for a primary r-process, 
while the dotted lines correspond to yields assuming a 
tr-process resulting from a collapse to a black hole assuming a LS220 EOS.  
%No Ba is 
%produced above 20 M$_\odot$ in this model.  Above
%20 M$_\odot$ Sr production depends on progenitor metallicity \citep{travaglio, raiteri}.
\label{yields}}
\end{figure}

A fascinating aspect of this work is the possibility that observed minima in [Ba/Fe], 
[Sr/Fe], and [Sr/Ba] may be directly related to the stiffness of
the nuclear EOS.  The lower limit of [Ba/Fe] as a function of
metallicity predicted in our model
can be reproduced in a GCE model assuming a contribution to r-process
elements from black holes produced in fallback supernovae.  
As shown in Figure \ref{sr_ba_fe_h}d, the softer the EOS, the lower the calculated [Ba/Fe] as 
a function of [Fe/H].
Thus the observed
lower limit in [Ba/Fe] as a function of metallicity appears to constrain the lower limit of the 
stiffness of the EOS.  
%While the current model is too primitive to predict an exact EOS,
%Future work
%will concentrate on refining these results.  
The tendency towards a soft EOS, however, is consistent
with prior experiments and observations \citep{exp,n-star mass}.  
While astronomical observations of neutron star masses
are able to predict a lower limit of the stiffness of the EOS, 
this model suggests a method of determining the upper limit.

Simulations also suggest that turbulence may explain the wide 
dispersion seen in these data sets, as the values of the predicted distributions do span more 
than the entire range of the data, and a more realistic model is necessary.
\acknowledgments
WA and TK were supported by the JSPS Grants-in-Aid for Scientific Research
(23224004 and 20244035) of the Ministry of Education, Culture, Sports,
Science and Technology of Japan.
%This work was supported through the NAOJ Visiting Professor Program.
RNB and MAF acknowledges support from the National Astronomical Observatory
and NSF grant \#PHY-1204486.

\clearpage

\clearpage
\end{document}